# Optimal scheduling of park-level integrated energy system considering ladder-type carbon trading mechanism and flexible load

Hongbin Sun [a], Xinmei Sun [a], Lei Kou [b], Benfa Zhang [c], Xiaodan Zhu [d]

a. School of Electrical Engineering, Changchun Institute of Technology, Changchun, China
b. Institute of Oceanographic Instrumentation, Qilu University of Technology (Shandong Academy of Sciences), Qingdao, China
c. State Grid Songyuan Power Supply Company, Songyuan, China
d. College of Mechanical and Electrical Engineering, Hulunbuir University, Hulunbuir Inmer Mongolia, China

Low carbon economy

a b s t r a c t

In an attempt to improve the utilization efficiency of multi-energy coupling in park-level integrated energy system (PIES), promote wind power consumption and reduce carbon emissions, a low-carbon economic operation optimization model of PIES integrating flexible load and carbon trading mechanism is constructed. Firstly, according to the characteristics of load response, the demand response is divided into four types: which can be shifted, transferred, reduced and replaced. Secondly, the PIES basic architecture is given by considering the combined heat and power generation coupling equipment, new energy and flexible load in the park. Finally, introducing the ladder-type carbon trading mechanism into the system and minimize the total operating cost, the low-carbon economic operation optimization model of PIES is established. The YALMIP toolbox and CPLEX solver are used to solve the example, the simulation results show that the participation of electrical and thermal coupled scheduling and flexible electric or thermal loads can significantly reduce the system operating cost, reduce the load peak-to-valley difference and relieve peak power consumption pressure.

## 1. Introduction

With the rapid development of economy and society, energy demand continues to grow, and problems such as energy crisis and environmental deterioration become increasingly severe (Ai and Hao, 2018). A park-level integrated energy system characterized by multi-energy interconnection, interworking and mutual economy can greatly improve the utilization efficiency of traditional energy and promote the utilization of renewable energy by coordinating and optimizing the production and conversion forms of various energy sources (Lu et al., 2021; Kou et al., 2020). It is one of the important carriers for realizing energy conservation and low carbon.

Most of the existing park-level integrated energy systems consider the overall economic cost, but seldom consider the additional environmental cost caused by carbon emissions. In order to reduce carbon emissions, the carbon trading mechanism can simultaneously improve the economy and environmental protection of the system (Fan et al., 2019). Considering the step carbon trading mechanism of the comprehensive energy system in the park, a dual-target operation optimization model with the minimum operation cost and the highest energy efficiency level is constructed (Ren et al., 2022). Initial carbon emission rights are allocated free of charge according to the actual output of nuclear power units, thermal power units and wind power units, and carbon trading costs are calculated by considering the actual carbon emissions of thermal power units, so as to effectively balance economic and low-carbon benefits (Li et al., 2019). A comprehensive power-heat-gas energy system containing waste treatment has been established, and a systematic low-carbon economy operation strategy based on carbon trading mechanism has been proposed (Yang et al., 2021). The stepped carbon trading was introduced into the economic dispatching model of regional integrated energy system, and the inhibition effect of the stepped carbon trading mechanism on carbon emission was proved (Qiu et al., 2022). These results provide a variety of effective methods for IES low-carbon modeling and carbon emission measurement, but they do not consider the impact of flexibility on system optimization operation (Wang et al., 2019). Considering the actual carbon emission of gas turbine and gas boiler, construct a carbon trading mechanism for integrated energy system. Finally, establish a low-carbon optimized operation model of integrated energy system based on the minimum objective function of energy purchase cost, carbon trading cost and operation and maintenance cost (Wei et al., 2022b). These results provide a variety of effective methods for PIES low-carbon modeling and carbon emission measurement, but none of them take into account the impact of load-side flexibility on system optimization operation.

Flexible load dispatching, as an important means to adjust the demand side energy performance of the power system, is widely concerned. Established more energy carrier production and transmission system and interconnected intelligent energy hub (smart energy hub, SEH) of more energy system optimization operation model, to minimize power cost and maximize user satisfaction as the goal, according to the system operators released energy price adjustment of the end user's energy consumption and internal operation plan, but only consider the characteristics can be reduced (Ni et al., 2018; Dolatabadi and Behnam, 2017; Kou et al., 2022). Introducing flexible load into the active distribution network realizes the reduction of system operation cost and network loss (Li et al., 2023; Jia et al., 2018). The microgrid dispatching model including the time-of-use electricity price mechanism and the flexible load is established. By planning the flexible load power consumption period, the electricity consumption efficiency and the clean energy consumption capacity of the distribution network are improved (Wang et al., 2017). An industrial IES operation architecture is constructed, and a demand response model is established considering the coupled response characteristics of cooling, heating and electricity demand (Zhao et al., 2022a,b). The above literature analyzed the ability of energy storage and controllable load to participate in integrated energy system operation regulation. However, for the interactive demand of integrated energy system, the synergistic advantage of load storage resources has not been fully played. Moreover, the response mechanism, form and restriction of controllable load are not considered enough, and the ''specific measures'' of regulated resources are not realized. In addition, the existing optimization model for storage and load side resources on the low-carbon operation of multi-energy systems focuses on the carbon emissions generated by primary energy consumption, and lacks the analysis of the flow and transfer of carbon emissions in the integrated energy system. Under the current low-carbon operation system of the integrated energy system, how to optimize all kinds of resources in the integrated energy system based on the existing architecture to achieve the goal of efficient energy conservation and emission reduction still needs in-depth study.

Most of the above studies on flexible load on the user side only consider one flexible load characteristic, while there are few studies on the controllability of heat load, and the load characteristic model is not perfect. In view of the above problems, this paper establish Shiftable, transferable, reducible and replaceable electric load model and extend them to thermal load. Takes PIES as the research object, introduce a carbon trading model, to analyze the regulation mechanism of flexible load in the park power network. Establish the cogeneration unit, electric heating equipment, electric-thermal energy storage and flexible load model in the electric-thermal coupling system, focusing on the resources of energy storage and flexible load. With the minimum sum of system operation cost and carbon emission cost as the objective function, a low-carbon environment coordinated operation model was constructed. Finally, the simulation is used to verify that the flexible load under the carbon trading mechanism can realize the economy and low carbon properties of peak shifting and valley filling and collaborative electric heating integrated energy system, and provide reference for the low-carbon economic operation of electric heating integrated energy system. The main contributions of this paper are as follows:

(1) The introduction of electrothermal flexible load, the establishment of the basic structure of the park-level comprehensive energy system with flexible load, while ensuring the economy, can improve the system's absorption capacity of new energy, especially wind energy;

(2) Compared with the traditional carbon trading pricing model, the stepwise carbon trading mechanism has a stronger binding force on carbon emissions and can play a better role in guiding carbon emissions reduction. Setting reasonable carbon trading parameters can play a role in guiding carbon emissions of the system and play a positive role in realizing the dual carbon goals;

(3) The carbon trading mechanism is introduced to establish the low-carbon economic operation optimization model of the park-level comprehensive energy system. It greatly improves the utilization efficiency of new energy, promotes the consumption of wind power and reduces carbon emissions.

Specific sections are assigned as follows. The first part introduces the overall architecture of the park-level integrated energy system and the mathematical model of the equipment in the park. The second part introduces the time-shifting characteristics of the flexible loads. The third part introduces a ladder-type carbon trading mechanism. The fourth part is the example simulation and analyzes the simulation results.

## 2. Bilateral structural frame of the park-level integrated energy system with coupled electric-heat flexible loads

The development of multi-energy complementary systems has brought great impetus to China's energy development and transformation and the construction of smart grids, which make the power system more intelligent and efficient. Because there are many devices in the park, each device has different parameters, and the operation situation is complicated, it is necessary to perform characteristic analysis and mathematical modeling of each device to complete the optimal operation of the entire system.

### 2.1. Bilateral optimal operation model based on the participation of flexible loads

In this paper, four main units constitute a park-type multienergy complementary system: the energy production unit, energy conversion unit, energy storage unit and energy consumption unit. Among them, energy production units include photovoltaics and wind power; energy conversion units include gas turbines, gas boilers, electric heating machines; energy storage units in this paper are mainly batteries and heat storage tanks; energy consumption units include electrical loads in the park and heat loads. The frame diagram of its energy flow is as Fig. 1.

The electric load and electric heating power consumption in the park are mainly provided by the upper-level power grid, photovoltaic power generation system, wind power generation system, gas turbine system and energy storage device. The upper natural gas network supplies gas for gas turbines and gas boilers.

New energy generally refers to renewable energy such as solar energy, biomass energy, and wind energy. New energy power generation technology is mainly developed through the exploration and utilization of new energy power generation, so its main cost is mainly the cost of fixed equipment. Therefore, in the integrated energy system, the operation and maintenance cost of system energy power generation is not considered. In addition, the new energy power generation technology uses clean energy and emits less pollutants, so new energy power generation should be used as much as possible (Li et al., 2018a,b). For photovoltaic power generation, wind power generation and other new energy power generation, the power generation cost is low considering the full consumption of clean energy, so its cost is not considered. This paper mainly considers the operating costs of controllable power sources such as gas turbines and energy storage units, the capacity electricity price for participating in the capacity ancillary service market, and the compensation electricity price for the load with demand response capability to participate in the dispatch of the integrated energy system .A gas turbine is a new type of thermal generator. Micro gas turbines have the characteristics of integrated expansion, convenient fuel, low energy consumption, etc. They are mainly used for distributed power generation, standby power stations, peak load power generation, etc. Among them, distributed power generation is widely used, such as urban commercial power generation, which can provide clean and reliable electric energy in regions, rural areas, etc., so gas turbines have good development prospects.

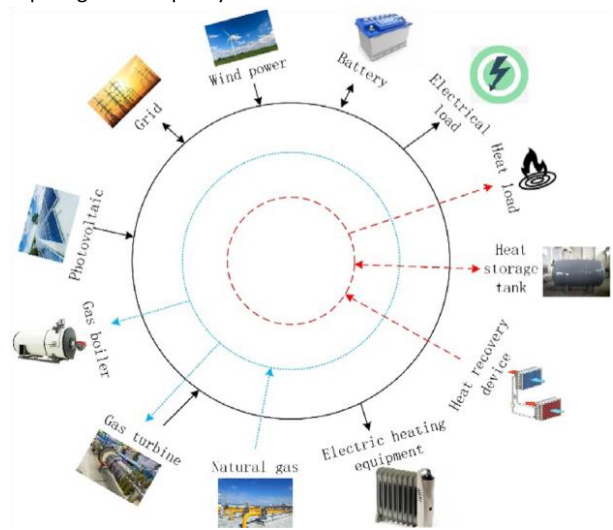

**Fig. 1.** Energy circulation diagram of the park-level integrated energy system.

*2.2. The park-level integrated energy system energy prediction model*

For photovoltaic power generation, wind power generation and other new energy power generation, the power generation cost is low considering the full consumption of clean energy. The new energy power generation technology uses clean energy and emits less pollutants, so new energy power generation should be used as much as possible.

*2.2.2. PIES energy conversion equipment mode*
  (1) Gas turbine
By consuming natural gas, the gas turbine can generate electric power to transmit electricity to the park and generate a large amount of waste heat. This waste heat can be used by waste heat boilers, which improves the utilization rate of thermal energy for the comprehensive energy system of the park.

*2.2.3. Energy storage devices model of the system optimize operation*
In the park-type multi-energy complementary system, there are usually various energy storage devices. In this paper, the storage battery (SB) is selected as the power storage device in the system. The state of charge (SOC) is usually used to describe the state of the energy storage device.

where $SOC$ is the state of charge of the battery; $P_d$ and $P_c$ are the discharge and charging power of the battery, respectively; $P_d^{max}$ and $P_c^{max}$ are the upper limit of the battery discharge power and charging power, respectively, and the upper limit is selected as the rated power that the device must plan; $C_{flag}$ and $D_{flag}$ are the battery charge and discharge, and the flags are 0 and 1 variables; $\eta_c$ and $\eta_d$ are the charging efficiency and discharging efficiency of the battery, respectively; $W_{ess}$ is the capacity of the battery; $SOC_{min}$ and $SOC_{max}$ are the lower and upper limits of the battery state of charge; $\Delta t$ is the operation optimization time. The battery has only one working state at each time, i.e., charging or discharging. The state of charge is generally 0.15–0.95. The heat storage tank is similar.

**3. Multi-type flexible load optimal dispatch based on the lowcarbon economy**

According to the forecast curve of load, wind turbine, and photovoltaic output in the park in the next 24 h, under the condition, that the system operation constraints are met the shift, translation, reduction, or replacement of flexible loads can be dispatched by rationally arranging the output of controllable units to assist coordination;auxiliary and coordinated energy storage equipment minimizes the total economic cost of PIES daily operation.

*3.1. Shiftable load*

The power supply time of the shiftable load can be changed as planned, the load must be shifted as a whole, and the power consumption time spans multiple scheduling periods. Assuming that the unit scheduling period is 1 h.

*3.2. Reduced load*

The load that can be cut can withstand a certain interruption or power reduction and reduce the time of operation, and it can be partially or completely cut according to the supply and demand situation. Unlike shiftable and transferable loads, shaving loads reduce the user's electricity consumption. A $\lambda$ variable of 0– 1 is used to represent $L_{cut}$ as the reduction status of the load that can be reduced in a certain period of time $\tau$; 0 implies that it is not reduced, and 1 implies $L_{cut}$ that it is reduced during period $\tau$.

*3.3. Replaceable loads*

For a certain type of heat load that can be directly supplied by heat energy or electric energy, electric energy is consumed when the electricity price is low, and heat energy can be directly consumed to satisfy the heat demand when the electricity price is high to mutually substitute electricity and heat energy.

Among the above loads, the power supply time of the shiftable load and transferable load can be changed as planned. The difference between the two is that the translational load requires the overall translation of the equipment. During the translation process, the equipment cannot be powered off, and the power supply time is fixed; i.e., the equipment has fixed power during the power consumption period before and after the translation, such as washing machines and disinfection cabinets. While the power of the power consumption period of the transferable load can be flexibly changed, the equipment is allowed to be interrupted during the power consumption period, and the power supply time does not have to be fixed; it must only satisfy the total demand of the load before and after the transfer. For example, the total charging capacity of the charging car is unchanged, and the charging time and charging power can be flexibly changed.

**4. Low-carbon optimization operation model of the pies based on carbon trading mechanism**

Considering the carbon trading mechanism and demand response, a multi-objective optimal operation model of the economy and energy efficiency of the integrated energy system in the park is constructed to improve the energy utilization and reduce carbon emissions.

*4.1. Ladder-type carbon trading model*

The carbon trading mechanism is to control carbon emissions by establishing legal carbon emission rights and allowing producers to trade carbon emission rights in the market. The regulatory department first allocates carbon emission rights quotas for each carbon emission source, and manufacturers combine their own quotas to perform reasonable production and emissions. If the actual carbon emissions are lower than the allocated quotas, the remaining quotas can be traded in the carbon trading market; otherwise, additional carbon emission quotas must be purchased. The stepped carbon trading mechanism model mainly consists of three parts: the carbon emission allowance model, the actual carbon emission model, and the stepped carbon emissions trading model (Chen et al., 2021).

There are three main types of carbon emission sources in the comprehensive energy system of the park: superior power purchase, gas turbines, and gas boilers. At present, the main quota method used in China is the free quota.

Compared with the traditional carbon trading pricing mechanism, to further limit carbon emissions, this paper adopts a tiered pricing mechanism. The stepped carbon trading mechanism divides multiple carbon emission right purchase intervals. Since the PIES must purchase more carbon emission allowances.

### 4.2. Objective function of the minimum operating cost and the minimum carbon transaction cost

The goal of the low-carbon economy optimization of the park's comprehensive energy system considering flexible loads is to reasonably arrange the output of each controllable unit of the park's IES and optimally schedule flexible loads if the constraints of each unit are satisfied.

where $F_{NET}$ is the interaction cost with the power grid; $F_{DG}$ is the cost of new energy power generation; $F_{MT}$ and $F_{GB}$ are the operating costs of the gas turbine and gas boiler, respectively; $F_{BAT}$ and $F_{HST}$ are the depreciation costs of the battery and heat storage tank, respectively; $F_L$ is the total compensation cost of dispatching flexible loads; $K_{NET}$ is the current electricity price; $P_{NET}(t)$ is the power exchanged with the grid, and the electricity purchase is positive; $K_{WT}$, $K_{PV}$, $K_{MT}$ and $K_{GB}$ are the operating cost coefficients of the fan, photovoltaic, gas turbine, and gas boiler, respectively; $P_{WT}$, $P_{PV}$, $P_{MT}$ and $P_{GB}$ are the output power of the fan, photovoltaic, gas turbine, and gas boiler, respectively; $K_{BAT}$ and $K_{HST}$ are the depreciation coefficients of the battery and heat storage tank, respectively; $P_{BAT}(t)$ and $P_{HST}(t)$ are the charging and discharging energy of the battery and heat storage tank, respectively, and the absorption is positive.

### 4.3. Operational constraints

(1) Load Balancing Constraints

The electric power and thermal power satisfy the following load balance constraints (Li et al., 2018a,b).

where $P_{WT/PV}$ and $P_{pred}$ are the output power and predicted output value of the WT/PV, respectively; $P_{MT/GB}$ and $P_N$ are the output power and equipment rated power of the gas turbine/gas boiler, respectively.

Other energy equipment constraints have been identified in the above model and are not repeated here.

(2) Energy storage equipment constraints:

The energy storage device constraints include energy storage capacity constraints, charging and discharging power constraints, charging and discharging state constraints, and constraints of zero full-cycle net exchange power:

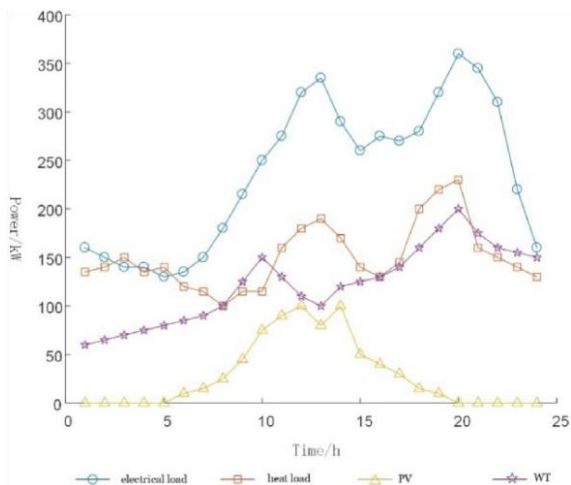

**Fig. 2.** Curves of load and PV/wind power output.

**Table 1** Time-of-use electricity price table (unit: ¥/kWh).

| Type | Peak period | Valley period | Normal period |
|---|---|---|---|
| Electricity purchase price | 0.82 | 0.25 | 0.53 |
| Electricity price | 0.65 | 0.22 | 0.45 |

where $E_{cap}(t)$ and $P_{ex}(t)$ are the energy storage capacity and charging and discharging power, respectively; $E_{cap,max}$, $E_{cap,min}$, $P_{ex,max}$, and $P_{ex,min}$ are the upper and lower limits of the energy storage capacity and charging and discharging power, respectively; $S(t)$ and $R(t)$ are the charging and discharging states, respectively, which are all 0–1 variables. Considering the service life of the battery, the upper limit of the number of times of daily charge and discharge is additionally limited to 8 times.

Aiming at the above model, the problem solved in this paper belongs to the mixed integer linear programming problem. First, the optimal operating cost of the park system is analyzed. Then, the lowest carbon transaction cost under the carbon trading mechanism is considered the objective function. Finally, after meeting various constraints, the MATLAB platform is used to call the CPLEX solver in the YALMIP toolbox to solve the model.

## 5. Case study

### 5.1. PIES basic data

This paper selects the PIES in Northeast China as the research object. The structure of the park mainly includes photovoltaics, fans, gas turbines, gas boilers, heat recovery systems, and energy storage equipment. We take the next day as a running cycle, take the time interval as 1 h, and divide it into 24 periods. The output forecast of photovoltaics and wind turbines and the forecast of electric heating load are shown in Fig. 2. The price of electricity purchased from the grid is shown in Table 1.

The flexible electrical/thermal load characteristics are shown in Table 2. Two types of shiftable electric loads are selected. Type 1 shiftable load start at 10:00 am and lasted for 4 h. The shiftable range time is 00:00–21:00. The type 2 shiftable load starts at 6:00 PM and lasted for 5 h. The shiftable range time is 00:00– 23:00. 17:00 pm and lasts for 4 h. The compensation price for all shiftable loads is 0.2 ¥ kWh$^{-1}$. Transferable electrical load power rang is 10–26.7 kW and continuous running time is 6 h. Transferable heat load power rang is 20–36.8 kW and continuous running time is 6 h. Transferable range and compensation cost for 6 h, 0.3 ¥ kWh$^{-1}$ respectively. For the reduced electric load, the sustainable reduction load time is 2–5 h, and the number of reduced load is 8. The reduced time range is 5:00 AM to 12:00 PM. The reduced heat load, the sustainable reduction load time is 3–8 h, and the number of reduced load is 10. The reduced time range is 11:00 AM to 17:00 PM. Alternative electrical loads type alternative power range is 10–20 kW and number of substitutions is 12. Alternative rang of alternative electrical loads is 7:00–19:00, and compensation cost 0.3 ¥ kWh$^{-1}$. Alternative heat loads type alternative power range is 10–33 kW and number of substitutions is 10. Alternative rang of alternative electrical loads is 12:00–24:00, and compensation cost 0.3 ¥ kWh$^{-1}$.

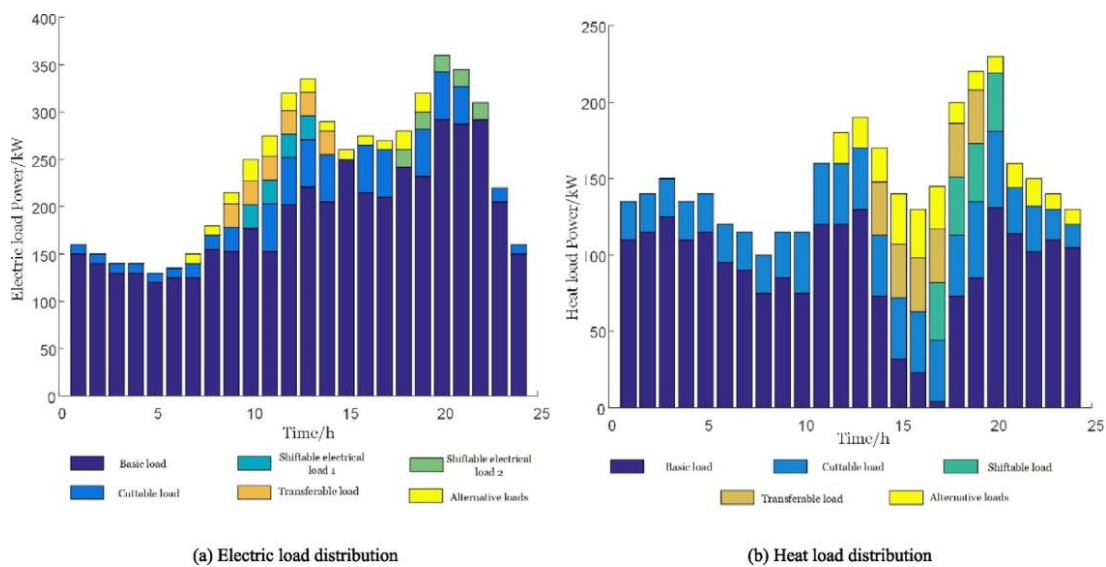

Fig. 3. User-side load distribution before system optimization.

**Table 2** User-side flexible load parameters.

| Load type | Start time/h | Duration/h | Shiftable range/h | Compensation Price/¥ kWh$^{-1}$ |
|---|---|---|---|---|
| Shiftable electrical load 1 | 10: 00 | 4 | [ 0:00~21:00] | 0.2 |
| Shiftable electrical load 2 | 18:00 | 5 | [ 0:00~23:00] | 0.2 |
| Shiftable heat load | 17:00 | 4 | [ 0:00~21:00] | 0.2 |
| Load type | Transfer power interval size/kW | Continuous running time/h | Transferable range/h | Compensation Price/¥ kWh$^{-1}$ |
| Transferable electrical load | [ 10 26.7] | 6 | [ 0:00~24:00] | 0.3 |
| Transferable heat load | [20 36.8] | 6 | [ 0:00~24:00] | 0.3 |
| Load type | Continuous cut time/h | Number of cuts/per | Cuttable range/h | Compensation Price/¥ kWh$^{-1}$ |
| Electric load can be reduced | [ 2 5 ] | 8 | [ 5:00~12:00] | 0.4 |
| Heat load can be reduced | [3 8] | 10 | [ 11:00~19:00] | 0.4 |
| Load type | Alternative power range/kW | Number of substitutions/per | Alternative interval/h | Compensation Price/¥ kWh$^{-1}$ |
| Alternative electrical loads | [ 10 20 ] | 12 | [ 7:00~19:00] | 0.3 |
| Alternative heat load | [10 33] | 10 | [ 12:00~24:00] | 0.3 |

**Table 3** Optimization results for different scenarios.

| Scenes | Total cost ¥/kWh | Carbon emission /kg | Peak-to-valley difference in heat load/kW | Electric load peak-to-valley difference/kW |
|---|---|---|---|---|
| 1 | 2946.4 | 4362.2 | 120 | 212 |
| 2 | 1583.6 | 2754.1 | 82 | 137 |
| 3 | 1476.5 | 2229.5 | 79 | 141 |
| 4 | 1006 | 1458.2 | 79 | 156 |

*5.2. Analysis of simulation results*

*5.2.1. Scenario comparative analysis*

To compare and analyze the economic and environmental benefits of the four scenarios, the IES scheduling model of the park in each scenario is solved, and the scheduling results are shown in Table 3.

The optimal scheduling results of the four scenarios in Table 3 show that the total cost of Scenario 2 is reduced by 46.3% compared with that of Scenario 1 after optimal scheduling. The reason is that the flexible scheduling capability of the flexible load is considered, which reduces the system cost. Due to the difference between peak and valley of the electric heating load, the park chooses a reasonable and more economical energy purchase time to reduce the cost; the cost of Scenario 3 is reduced by 6% based on Scenario 2 because the replaceable loads are considered in Scenario 3, and the electric heat loads can be replaced by one another. The replaceable load can convert part of the electrical load into thermal load during the energy purchase peak period and convert the electrical load into thermal load during the energy purchase valley period, which further reduces the peak period electrical load and energy purchase cost. Scenario 4 is compared with Scenario 1. The total cost has dropped by 65.9%, and the carbon emission has dropped by 67%. This is the result of the coordinated optimization of flexible load and carbon trading. The introduction of the carbon trading mechanism allows the park to have basic carbon emission rights, which can offset part of the cost of carbon emissions to reduce costs. Scenario 4 is compared with Scenarios 2 and 3.

The cost of Scenario 3 is 3 1.9% higher than that of Scenario 4, and the cost of Scenario 2 is 3 6.5% higher than that of Scenario 4, since Scenarios 2 and 3 do not introduce a carbon trading mechanism. The full cost of purchasing all carbon emissions is required.

*5.2.2. Influence of the flexible load on the optimal operation*

Scenario 1 does not consider the flexible load to participate in the system operation. The output diagram of each energy source is shown in Fig. 4, and the electric thermal load distribution diagram is shown in Fig. 3.

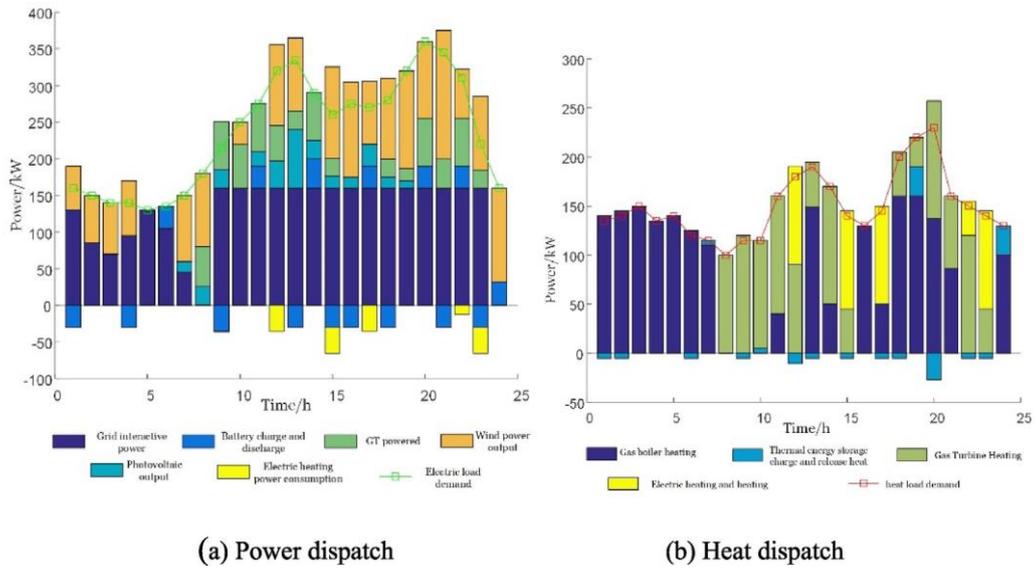

(a) Power dispatch                     (b) Heat dispatch

**Fig. 4.** Output of each energy source after system optimization in Scenario 1.

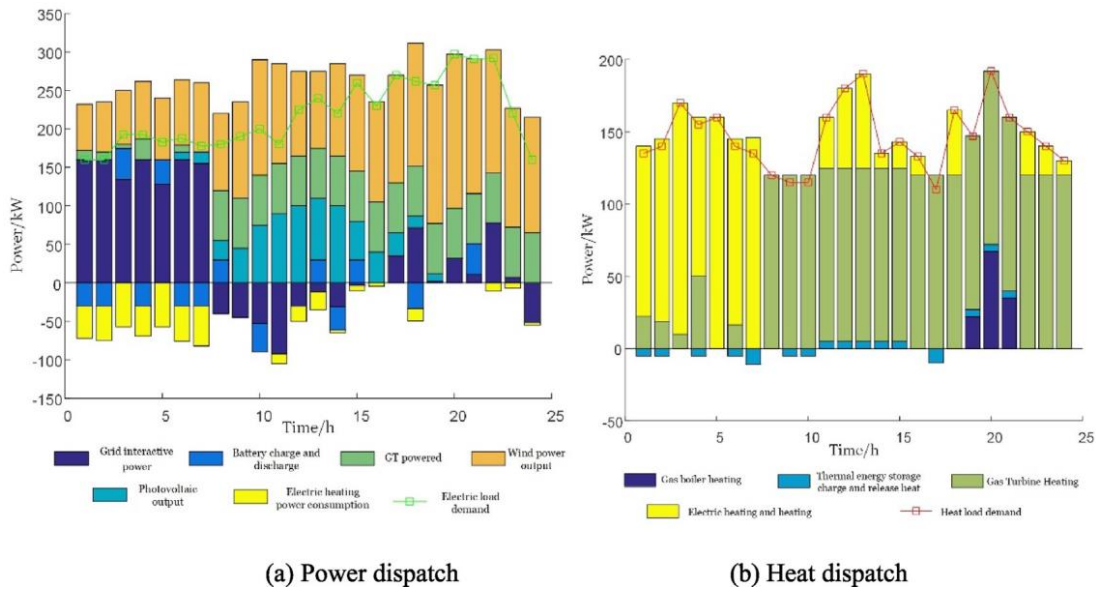

(a) Power dispatch                     (b) Heat dispatch

**Fig. 5.** Output of each energy source after system optimization in Scenario 2.

Figs. 4 and 5 show the output of each energy source after the optimal operation of the system in Scenarios 1 and 2. From Fig. 4(a), in Scenario 1, which does not consider the addition of any flexible load, the electricity consumption behavior of users is mostly concentrated in 11: 00–16: 00 and 19: 00–22: 00. These two periods are also peak periods of electricity prices. In the electricity price valley period of 00: 00–00: 08, there is less power purchase from the grid, and the phenomenon of wind curtailment in Scenario 1 is serious, which increases the operating cost of the system. In Scenario 2, the flexible load is added to the system for optimal operation. Fig. 3(a) of the load distribution on the user side before optimization shows that for the electric load in the period of 00: 00–08: 00, the required electric load is in the low-level period, the electricity price during this period also belongs to the trough period, and the electricity purchase price is 0.25 ¥/kWh. The analysis of the energy output of Scenario 2 in Fig. 5(a), shows that the photovoltaic units have no output during the period of 00: 00–08: 00. At this time, the operation and maintenance cost of wind turbines is 0.52 ¥/kWh, which is higher than the electricity purchase price, so the grid power supply is preferred, and the photovoltaic units have no output before 00:09. At this time, the gas turbine is in the mode of constant electricity by heat. Fig. 3(b) shows that the total amount of heat load in the period of 00:00–08:00 is higher than the total amount of electric load in this period, so excess electricity will be produced, and excess electricity will be generated. The electricity is only enough to charge the battery; electricity prices are higher during the two

00, etc.), the gas turbine provides heat, and the extra heat is collected by the thermal energy storage to release heat during the peak heat load period, and the thermal energy storage helps shave peaks and fill valleys in the system.

Based on the output of each energy source after system optimization in Figs. 4 and 5 and the load distribution after optimization in Figs. 3 and 6, the addition of flexible loads can effectively reduce the total operating cost and load peak-to-valley difference of the integrated energy system.

Figs. 6 and 7 show the load distribution after the optimized operation of the system in Scenario 2 and Scenario 3, respectively. The new energy and gas turbines in these two scenarios have basically identical outputs, which results in different costs and carbon emissions. The main reason is that compared with Scenario 2, Scenario 3 adds electric and heat replaceable loads based on existing flexible loads (including shiftable loads, cuttable loads, and transferable loads).

A comparison of Fig. 6 with Fig. 3 and our analysis show that for a flexible electrical load, during the period of 19: 00– 21: 00, 16: 00–17: 00, and 11: 00–14: 00, electric load reduction occurs; this period is also the peak period of electricity price. The cutoff characteristics of the power load that can be cut are shown in Table 2. Shiftable load 1 is translated from 11:00–14:00 to 03:00–06:00, shiftable load 2 is translated from 18:00–22:00 to 03:00–07:00, and the translational load as a whole is translated from the peak period of the electricity price to the valley segment of the electricity price. The translation result also conforms to periods of 11:00–15:00 and 19:00–22:00, and new energy is abundant, so wind and photovoltaic resources can be fully utilized to save system operation costs.

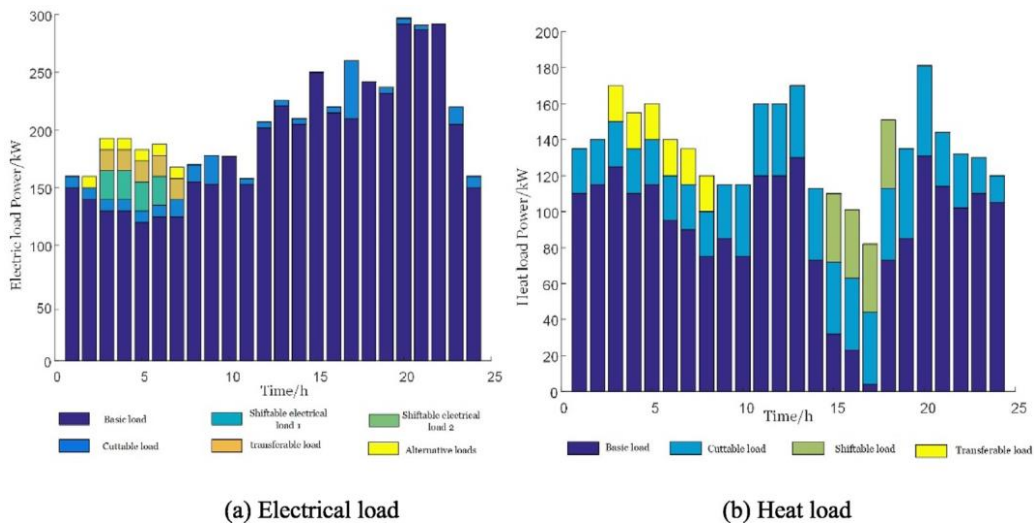

**Fig. 6.** Distribution of various types of loads on the user side after system optimization in Scenario 2.

For heat loads, most of the heat is provided by gas turbines and supplemented by thermal storage tanks, electric heating equipment and heat recovery systems. Fig. 5(b) shows that the 00:00– 00:08 period gas boilers and heat storage tank of heat supply reduces the heat supply for gas turbines, and the heat supply is reduced by the working mode of the gas turbine in terms of heat and electricity. In addition, the power generation is reduced, and the reduction of gas turbine power generation affects power dispatching; thus, the system can purchase more power from the grid when the electricity price is low, and the excess power is absorbed by the battery, which reduces the operating cost of power dispatching during this period. From Fig. 6(a), the two periods of 11: 00–16: 00 and 19: 00–22: 00 are the peak electricity consumption period and peak electricity price period, and the full power of the gas turbine can simultaneously compensate for the shortage of electricity. From Fig. 6(b), during the low-heat-load periods (00: 0o–00:02, 08: 00–10: 00, 14: 00–17:

the characteristics of the translational load. The transferable load is transferred from 09:00–14:00 to 02:00–07:00. The specific characteristics of the shiftable load and transferable load are shown in Table 2. From the analysis results, we can conclude that the total number of time periods for shiftable load 1 and shiftable load 2 does not change during the translation process, and the duration of transfer and total power consumption do not change either. Although there is no change, the total power consumption of the transfer has been reduced, so the transferable load is more flexible than the shiftable load. A comparison of Fig. 7 with Fig. 3, i.e., Scenario 3 with Scenario 2 shows that the scheduling situation of transferable, shiftable and cuttable loads in Scenario 3 is basically similar to that in Scenario 2. Before optimization, the replaceable electric load is evenly distributed in 13 time periods of 08: 00–20: 00; after optimization, in the electricity price peak periods of 12: 00–14: 00 and 18: 00–19:00, the replaceable electric load is replaced by a thermal load. Similarly, part of the thermal load is replaced by the electric load during the electricity price valley period of 00:00–08:00. By cooperating with various flexible loads, the load curve is smoother. The goal of shaving peaks and filling valleys has been achieved, which further eases the pressure on peak electricity consumption.

Through the above comparative analysis, the addition of flexible loads to participate in the optimal scheduling of the system in Scenarios 2 and 3 can effectively reduce the peak-to-valley difference of loads. The collaborative participation of multiple types of loads can fully exploit the advantages of multi-energy complementarity and flexibly realize the source-load interaction.

#### 5.2.3. The combined effect of carbon trading mechanism and flexible loads

To verify the impact of the introduction of carbon trading on the IES, the dispatch results of Scenario 3 and Scenario 4 are analyzed, and the output of each energy source in the scenarios before and after the introduction of carbon trading can be obtained, as shown in Fig. 8.

Scenario 4 considers the flexible load and carbon trading mechanism to participate in the system optimization operation. The scheduling results are shown in Fig. 9. Fig. 9(b) shows that the gas turbine is always in full power supply for heat supply. During the peak heat period, electric heating equipment is used to supplement the heat, and the excess heat is collected in the thermal energy storage equipment. When the heat supply is insufficient, supplementary heat is conducted, and the total number of cycles of thermal energy storage is 8. Comparing Figs. 8 and 9, the gas turbine works in the mode of electricity by heat. The gas turbine is fully powered so that the system reduces the power purchase behavior of the grid during the peak period of electricity and can sell more electricity to the grid or store it in the battery. Reducing the electricity purchase behavior is equivalent to reducing carbon emissions, which can sell excess carbon allowances to the grid, improve environmental benefits, and reduce operating costs. Fig. 9(a) shows that when electricity is used during peak hours 11:00–14:00, the system can reversely transmit the excess electric energy to the grid under the condition of completely consuming new energy, and the power supply in the entire period of Scenario 4 is mostly powered by new energy. The consumption of energy reaches the maximum, which saves the operation and maintenance cost of the system while protecting the environment. The batteries in both Scenarios 3 and 4 play the role of shaving peaks and filling valleys, while Scenario 4 is more inclined to use new energy for battery charge and reasonably use the battery to supply power to the user's load during the peak period of electricity consumption.

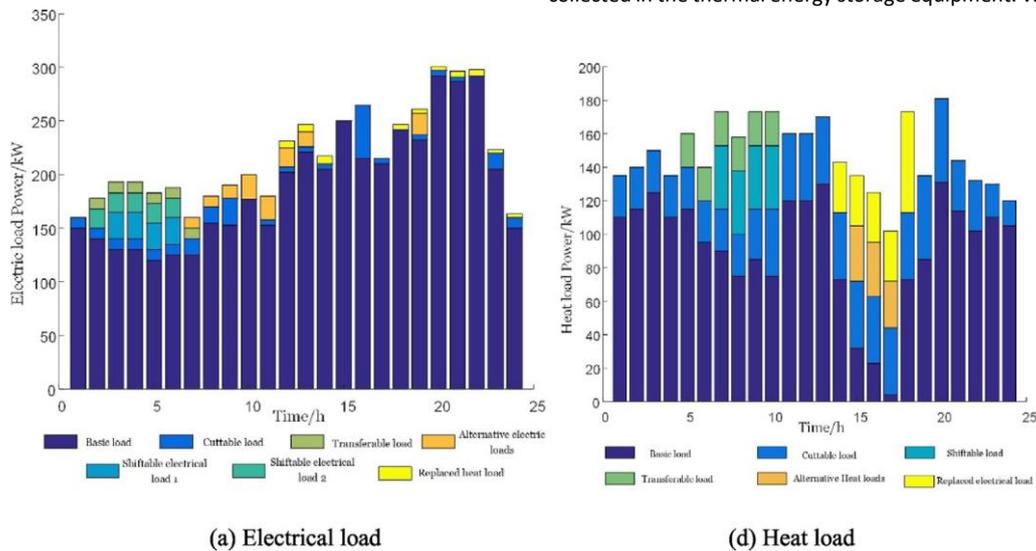

Fig. 7. Distribution of various types of loads on the user side after system optimization in scenario 3.

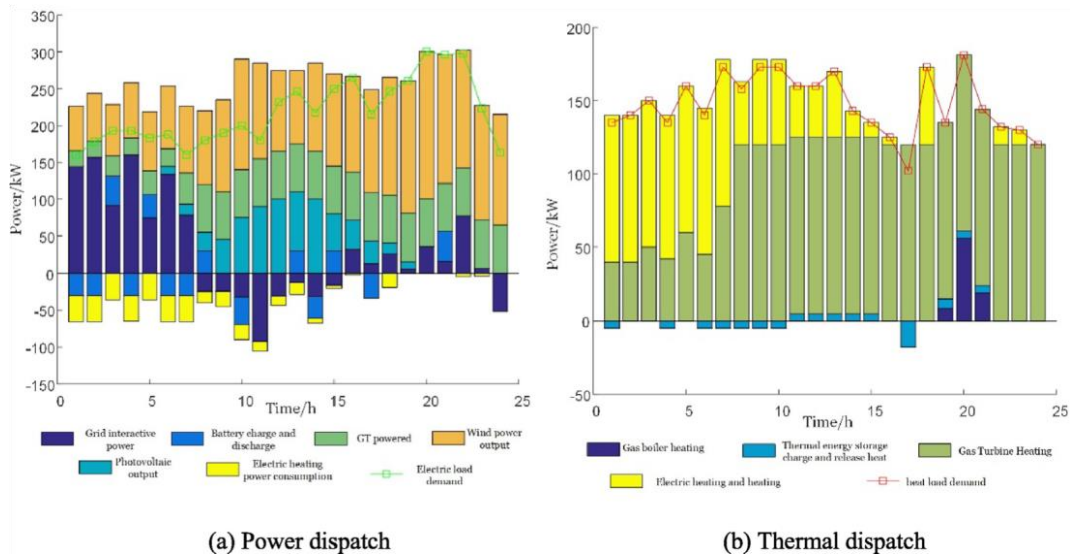

Fig. 8. Output of each energy source after system optimization in Scenario 3.

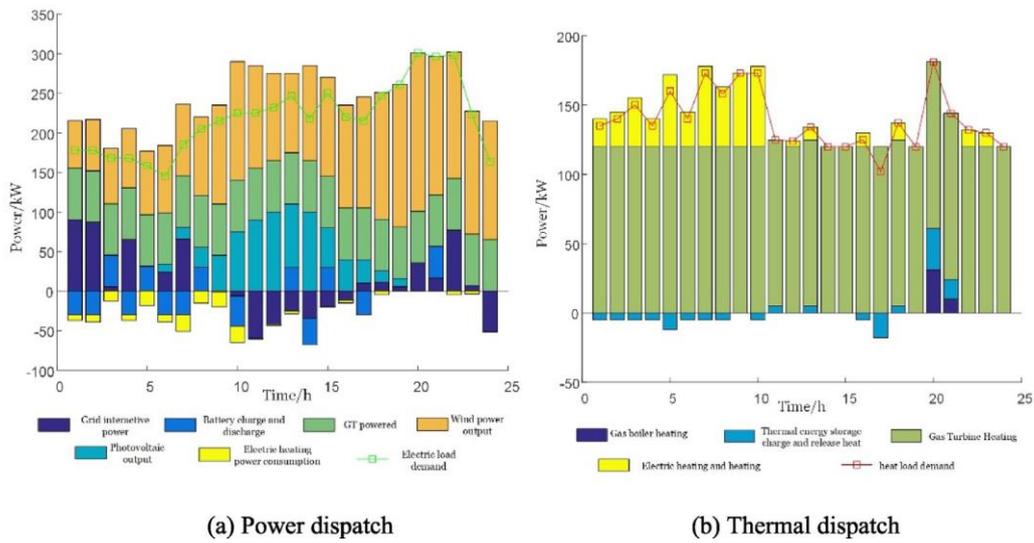

Fig. 9. Output of each energy source after system optimization in scenario 4.

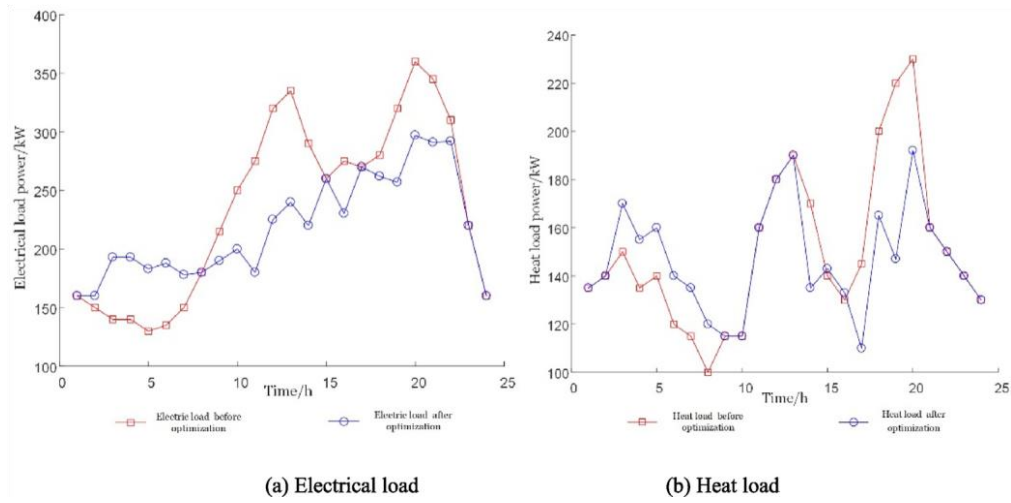

Fig. A.1. Comparison of load distribution curves before and after system optimization in scenario 2.

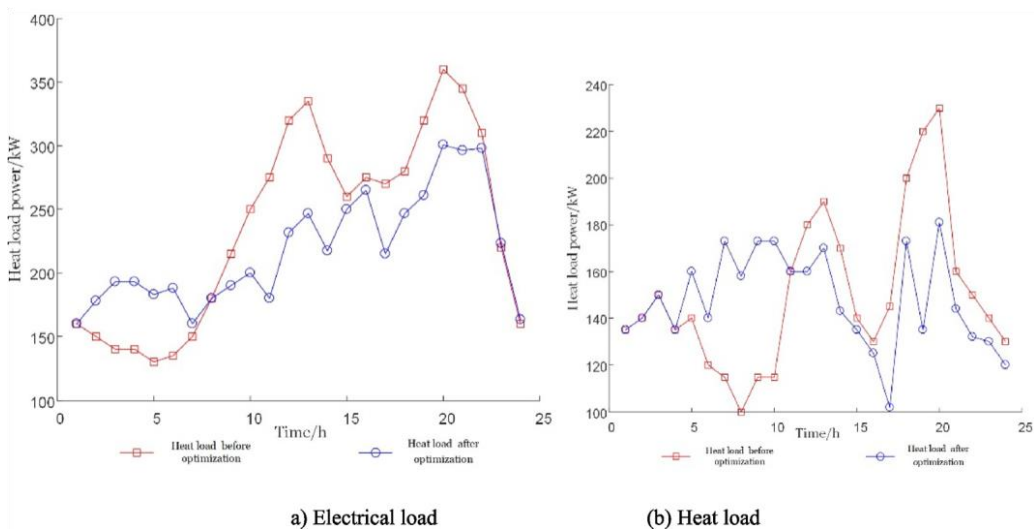

Fig. A.2. Comparison of the load distribution curves before and after system optimization in scenario 3.

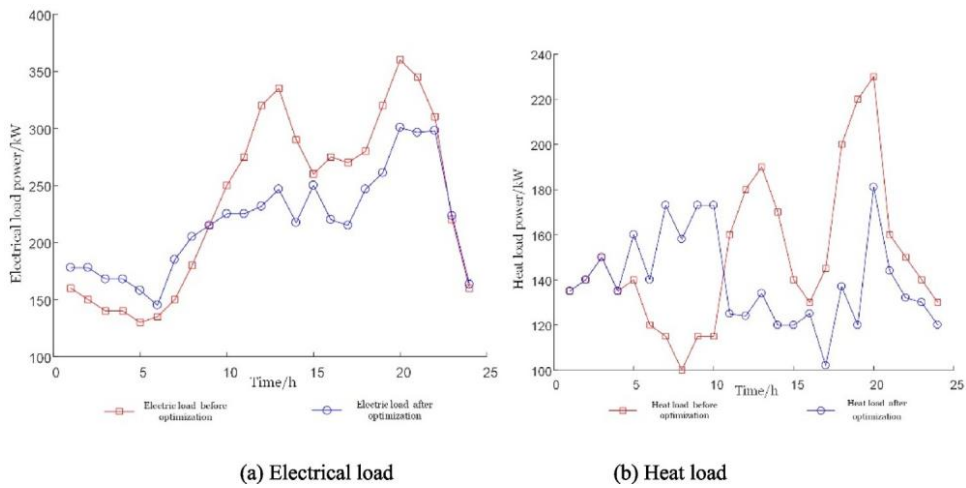

**Fig. A.3.** Comparison of the load distribution curves before and after system optimization in scenario 4.

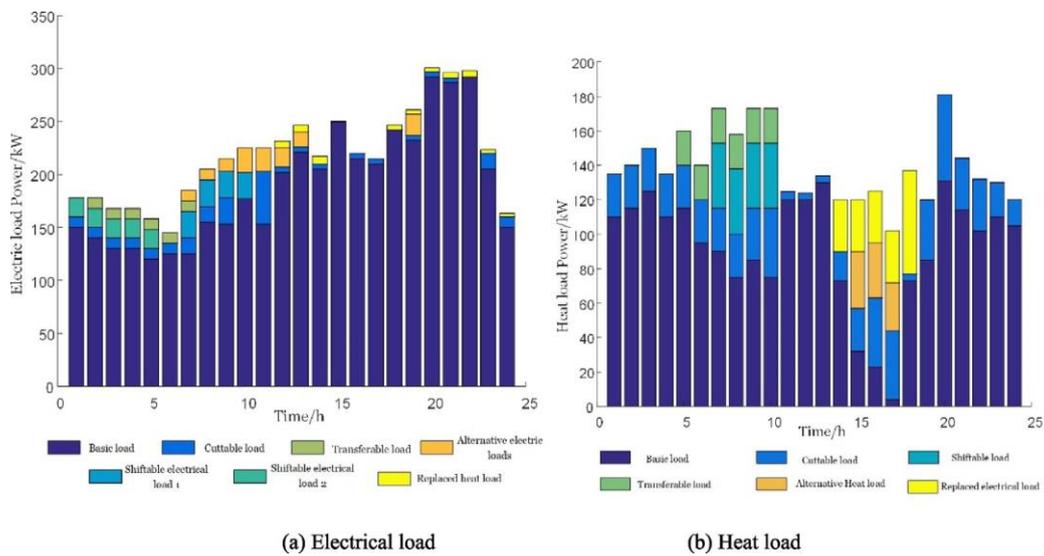

**Fig. A.4.** Scenario 4 Load distribution after the system optimized operation.

From the above analysis, the carbon trading mechanism can reasonably guide the development of the park system toward low-carbon development and optimize the structure of the park's comprehensive energy system. The participation of flexible loads can enable the system to flexibly dispatch the output of various energy sources and achieve a benign interaction between source and load. The synergistic optimization of the two can comprehensively consider the environmental protection and economy of the integrated energy system based on their respective advantages and realize the low-carbon economic operation goal of the park.

## 6. Conclusion

In this paper, for the park level comprehensive energy system introducing wind power, photovoltaic and other new energy, on the basis of the basic system architecture, the impact of the power/ heat flexible load on the optimal operation of the park level comprehensive energy system is revealed; The laddertype carbon trading mechanism is introduced into the system operation optimization, and a low-carbon economic optimization model of PIES is constructed. Compared with the scheduling and optimization results of the park-level integrated energy system in four different scenarios, the following conclusions are obtained based on simulation results:

(1) Fully consider the impact of electric thermal soft load on demand side load adjustment, can guide users timely adjustment of energy use time, to achieve peak load shifting
and valley filling, effectively improve the system economy, environmental protection and the utilization rate of new energy.

(2) Compared with the traditional carbon trading model, the stepped carbon trading mechanism is more binding on carbon emissions, can better guide the effect of emission reduction, give full play to the mobilization ability of the carbon trading market, and solve the problem of environmental pollution.

(3) Under the carbon trading mechanism, considering the flexible load not only transfers part of the load in the high price period to the low price period and reduces part of the load energy use, but also realizes the mutual substitution of electric energy and heat energy on the user side and smooth the load curve; At the same time, the flexible choice of energy purchase methods can effectively coordinate the operation economy and low carbon performance of the system.

The subsequent research should further consider the practical characteristics such as the source and load uncertainty and the equipment variable working conditions, in order to obtain a more perfect mathematical model and scheduling strategy.